\documentclass[amsmath,amssymb,iopams]{iopart}

\usepackage{graphicx}

\begin{document}

\title{Finite-Size Bosonization and Self-Consistent Harmonic Approximation}
\author{C.\ Mocanu, M.\ Dzierzawa, P.\ Schwab and U.\ Eckern}
\address{Institut f\"ur Physik, Universit\"at Augsburg, D-86135 Augsburg, Germany}
\
\ead{carmen.mocanu@physik.uni-augsburg.de}
\date{\today}

\begin{abstract}
The self-consistent harmonic approximation is extended in order
to account for the existence of Klein factors in bosonized Hamiltonians.
This is important for the study of finite systems where Klein factors
cannot be ignored a priori.
As a test we apply the method to interacting spinless fermions with
modulated hopping. We calculate the finite-size corrections to the
energy gap and the Drude weight and compare our results with
the exact solution for special values of the model parameters.
\end{abstract}


\section{Introduction}
Bosonization techniques have been extensively used in the study of one-dimensional
electron and spin systems \cite{Haldane81,Delft98,schulz00}.
The success of the method is based on the fact that the low energy
properties of fermions are determined by the states close to the Fermi
surface which in one dimension consists only of two points.
When the spectrum is linearized around the Fermi points
the Hamiltonian can be expressed in terms
of bosonic operators associated with
particle-hole excitations of momentum $q$.
In addition, scattering processes
between left and right Fermi points require to introduce extra operators,
the so-called Klein factors.

Perturbations like impurity scattering or a modulation of the hopping
lead to nonlinearities in the bosonized Hamiltonian, for which an exact solution is known
only in some special cases \cite{baxter1982}.
In general one has to resort to approximative methods like
renormalization group calculations.
Another more intuitive method is the self-consistent harmonic approximation (SCHA)
where the non-linear terms are replaced by a harmonic potential
with parameters to be determined  self-consistently according to a variational principle for the
energy.

The SCHA has been successfully applied to various non-linear field theories
\cite{Coleman75,Nakano81,Fukuyama,gogolin1993,rojas1996,gouvea1999,Cosima}.
In the context of bosonized Hamiltonians, however,
the existence of Klein factors has been ignored in these approaches.
This may be justified for infinite systems \cite{schulz00},
but in general the Klein factors have to be treated carefully as pointed
out for example in the context of impurity models and two-leg ladders
\cite{Delft98,Kotliar96,schonhammer01,fjarestad02}.
In the following we present an extension of the SCHA which
treats the bosonic fields and the Klein factors on equal footing.
As a prototype model for this study we consider the lattice model of spinless fermions at half filling
including nearest-neighbor interaction and a modulation of the hopping due to a periodic lattice distortion.

In the next section the Hamiltonian of this model is bosonized and a trial
Hamiltonian is constructed which serves as the basis for the SCHA.
In Sect.\ 3 we calculate the energy gap and the Drude weight and compare our results
with exact results that are available for certain values of the model parameters.
Finally we conclude with a brief summary in Sect.\ 4.
\section{Model and formalism}
We consider a one-dimensional model of spinless fermions with static
dimerization
\begin{eqnarray}
\label{eq:dim-term}
   H & = & -t\sum_{i}(1+(-1)^i u)(c^+_ic_{i+1} + c^+_{i+1}c_i )\nonumber\\
     & &{} + V \sum_{i}n_i n_{i+1}
\end{eqnarray}
where $u$ is the dimerization parameter that leads to a periodic
modulation of the hopping amplitude $t$, and $V$ is the strength of the
nearest-neighbor interaction. At half filling and in the range $-2< V/t < 2 $ the bosonized Hamiltonian of
this model reads \cite{Delft98}
\begin{eqnarray}
\label{eq:totalhamilt}
 H & = &\int_0^L \frac{{\rm d} x}{2\pi}\left\{\frac{v}{g} \, {:} (\partial_x\phi)^2 {:} \,
 + v g \, {:} (\partial_x\theta)^2{:} \, \right\}+ \frac{\pi v}{2Lg}N^2
  \nonumber\\
  & &
 +  \frac{\pi vg}{2L}J^2- \frac{itu}{\pi a}\int_0^L {\rm d}x \ F_L^+F_R  \ {\rm e}^{2i\phi(x)}
  + h.c.
\end{eqnarray}
where $\phi$ and $\theta$ are conjugate fields.
The renormalized Fermi velocity $v = \pi t\sin(2\eta)/(\pi-2\eta)$ and the
Luttinger parameter $g = \pi/4\eta$ are related to the interaction
according to $V=-2t\cos(2\eta)$.
The normal ordering operation ($: \quad:$) is introduced to
avoid the divergences associated with unphysical states assumed by
the Luttinger model.
The parameter $a$ is needed in the bosonization formalism 
to remove ultra-violet divergences in certain $k$-summations, and appears in Eq.~(\ref{eq:totalhamilt})
since the exponential $\exp( 2  i \phi(x) )$ is not normal ordered. For a lattice model,
$1/a$ can be interpreted as a kind of effective band-width \cite{Delft98},
whereas in the continuum the limit $a \rightarrow 0 $ has to be performed in order
to obtain the correct $\delta$-function anticommutators of the fermionic field operators \cite{Haldane81}.
$N$ and $J$ are the
charge and current operators, defined as $N=N_L+N_R$ and
$J=N_L-N_R$, respectively, where $N_L$ and
$N_R$ count the number of left
and right moving particles with respect to the filled Fermi sea. The Klein factors
$F^+_{L/R}$ and $F_{L/R}$  rise or lower the
number of left and right moving fermions by one (which no combination of bosonic operators
can ever do), and they assure also
that fermions of the two different species anticommute.
The dimerization, described by the non-linear term of the Hamiltonian,
breaks the conservation of the current $J$ and is responsible for the opening
of an energy gap.

To discuss the gap formation quantitatively, we use the self-consistent
harmonic approximation (SCHA) which has originally been introduced for
the sine-Gordon model  \cite{Coleman75}.
The idea is to construct an exactly solvable trial Hamiltonian $H_{\rm tr}$ where the bosonic
fields in (\ref{eq:totalhamilt}) are decoupled from the Klein factors and where the non-linear
terms $\sim {\rm e}^{\pm2i\phi}$ are replaced with a quadratic form $\sim \phi^2$.
Accordingly we chose $H_{\rm tr} = H_{\rm tr}^\Delta + H_{\rm tr}^B$ as the sum of two commuting parts
\begin{equation}
\label{eq:HtrialA}
H_{\rm tr}^\Delta = \int_0^L \frac{\rm{d}x}{2\pi}
  \left\{\frac{v}{g}(\partial_x\phi)^2 + vg(\partial_x\theta)^2
  + \frac{\Delta^2}{v g} \phi^2(x)\right\}
\end{equation}
and
\begin{equation}
\label{eq:HtrialB}
H_{\rm tr}^B = -iB(F^+_LF_R-F^+_RF_L) + \frac{\pi vg}{2L}J^2
\end{equation}
each of them depending on a single variational parameter, $\Delta$ and $B$, respectively.
Notice that from the standard trial Hamiltonian alone (Eq.~(\ref{eq:HtrialA})), only
the trivial solution $\Delta =0$ can be obtained.
Therefore the non-standard term Eq.~(\ref{eq:HtrialB}) is necessarily required to
describe the phase transition in the system.
Since we are only interested in properties at zero temperature we use the
normalized ground state $|\psi_0\rangle$ of $H_{\rm tr}$ as a variational
wave-function for the Hamiltonian (\ref{eq:totalhamilt}). This provides us
with an upper bound $\tilde E$ for the ground state energy $E$ of (\ref{eq:totalhamilt})
due to the inequality
\begin{equation}
\label{eq:variational}
E\le\tilde{E}=\langle \psi_0 |H  |\psi_0\rangle
. \end{equation}
Minimizing $\tilde E$ with respect to the variational parameters $\Delta$ and $B$
yields the following set of equations:
\begin{equation}
\label{eq:varpar}
\begin{array}{lll}
   \Delta^2  & = & -(4vg {tu}/{a}) E_0'(B) {\rm e}^{-2\langle\phi^2\rangle_{\rm tr}} \\[2mm]
      B      & = & (tu /{\pi a}) L {\rm e}^{-2\langle\phi^2\rangle_{\rm  tr}}
\end{array}
\end{equation}
where $E_0(B)$ is the ground state energy of $H_{\rm tr}^B$ and $E_0'(B)=dE_0(B)/dB$.
Since $H_{\rm tr}^\Delta$ is bilinear in the field operators
it is straightforward to calculate the equal time correlation
function of the phase field $\phi(x)$ entering Eq.\ (\ref{eq:varpar}):
\begin{equation}
\label{eq:phi2}
\langle\phi^2\rangle_{\rm tr} = \frac{\pi v g}{L}
\sum_{k>0} \frac{{\rm e}^{-k a}}{\sqrt{v^2 k^2 + \Delta^2}}
\end{equation}
where the $k$-values in the sum are multiples of $2\pi/L$.
\section{Energy gap and Drude weight}
In order to solve the set of equations (\ref{eq:varpar}) we have to
calculate the ground state energy $E_0(B)$ of the trial Hamiltonian $H_{\rm tr}^B$.
The products of Klein factors $F^+_L F_R$ and $F^+_R F_L$
appearing in $H_{\rm tr}^B$ change the quantum number of the current operator
from $J$ to $J \pm 2$. $H_{\rm tr}^B$ can therefore be regarded
as a particle moving in a harmonic potential along the $J$-axis.
For $L \rightarrow \infty$ the potential energy term in $H_{\rm tr}^B$ vanishes
and the kinetic energy term with imaginary
hopping $i B$ yields the ground state energy $E_0(B) = - 2 B$.
In this limit we reproduce the results of Ref.\ \cite{Cosima}, where
the Klein factors have been ignored.
Replacing the sum in Eq.\ (\ref{eq:phi2}) by an integral
we obtain
\begin{equation}
\label{eq:phi2a}
\langle\phi^2\rangle_{\rm tr} = \frac{g}{2} \ln{\frac{\Delta_0}{\Delta}}
\end{equation}
where $\Delta_0 = 2 v {\rm e}^{-\gamma} / a$, $\gamma=0.5772$ is Euler's constant,
and $\Delta \ll \Delta_0$ has been assumed.
Inserting Eq.\ (\ref{eq:phi2a}) into Eq.\ (\ref{eq:varpar}) and using
$E_0'(B) = -2$ yields
\begin{equation}
\label{eq:delta}
\frac{\Delta}{\Delta_0} = \left(\frac{u}{u_0}\right)^{1/(2- g)}
\end{equation}
with $u_0 = {\rm e}^{-\gamma} \Delta_0 / 4gt$. For $g > 2$
Eq.\ (\ref{eq:delta}) diverges as $u \rightarrow 0$ and there is only
the trivial solution $\Delta = 0$, i.e.\ the line $g = 2$
marks the transition from a gapless to a gapped phase
in the $g$-$u$ plane.
In the spinless fermion model
this corresponds to the line $V/t = - \sqrt{2}$ in the $V$-$u$ phase diagram.
The value $g_c = 2$ obtained within the variational approach is in accordance
with the exact result based on a mapping of the bosonized Hamiltonian
on the Ashkin-Teller model \cite{Kohmoto81} for $u \rightarrow 0$.
However, according to renormalization group calculations \cite{Zang95} and confirmed
numerically using density matrix renormalization group \cite{Cosima}, the phase boundary
in the $g$-$u$ plane shifts to larger values of $g$ with
increasing $u$ while the SCHA gives a vertical line.
On the other hand,
the exponent $1/(2 - g)$ characterizing the opening of the gap
is exact \cite{Nakano81}. In particular for non-interacting fermions $(V = 0)$
the Luttinger liquid parameter is $g = 1$ and therefore
$\Delta \propto u$ in agreement with the exact energy gap $\Delta_{\rm ex} = 4 t u $.
At $V = 2t$, where Umklapp scattering becomes relevant and leads to the
opening of a correlation gap, Eq. (\ref{eq:delta})
yields $\Delta \propto u^{2/3}$ which agrees with the exact result
$\Delta_{\rm ex}\propto u^{2/3}/ \sqrt{|\ln u|}$ up to the
logarithmic correction \cite{Uhrig}.

In the following we consider a system of finite length $L$.
In order to calculate the ground state energy of
$H_{\rm tr}^B$ it is convenient to
switch to the momentum representation, $J \rightarrow
-i {\rm d}/{\rm d}p_J $, where the trial Hamiltonian reads
\begin{equation}
\label{eq:HB}
 H_{\rm tr}^B=-\frac{\pi vg}{2L}\frac{d^2}{dp_J^2} +2B \cos(2p_J)
\end{equation}
with periodic boundary conditions for the wave-function,
$\Psi(p_J+\pi)=\Psi(p_J)$.
The corresponding Schr\"odinger equation is known as Mathieu's equation.
Accordingly, the ground state energy of $H_{\rm tr}^B$ has the following
asymptotical behavior \cite{Abramowitz}
\begin{equation}
\label{eq:E0}
E_0(B) \approx
\left\{
\begin{array}{cc}
-{B^2L}/({\pi vg})  & \textrm{for } LB \ll v\\
-2B
& \textrm{for } LB \gg v
\end{array}
\right.
\end{equation}
Combining the two equations (\ref{eq:varpar}) one obtains
\begin{equation}
\label{eq:delta2}
\Delta^2 L^2 = - 4\pi v g E_0'(B) B L
\end{equation}
which can be used to relate the two limiting cases of Eq.\ (\ref{eq:E0}) with
two different physical situations:
The condition $LB \gg v$ is equivalent with $L \Delta \gg v$ and therefore
the results for $\langle\phi^2\rangle_{\rm tr}$
and $\Delta(u)$ are the same as for the infinite system i.e.\ the energy gap
depends algebraically on the dimerization.
On the other hand, the condition $LB \ll v$ implies that $L\Delta \ll v$.
In this case the energy gap is purely due to the finite system size and of
order $2\pi v/L$.
Thus the crossover from a finite-size gap to a true dimerization gap 
coincides with the crossover between the regions where Klein factors
are relevant or can be ignored.

In general, the set of equations (\ref{eq:varpar}) can only be solved
numerically. Fig.\ 1 shows the size-dependent energy gap $\Delta(L) = \sqrt{(2\pi v/L)^2 + \Delta^2}$,
i.e.\ the lowest excitation energy of the bosonic modes,
as function of $u$ for several values of the system size $L$; notice that  we consider a system with periodic
boundary conditions, where the smallest wave number is $2 \pi/L$.
The Luttinger parameter is $g=3/4$  which corresponds to $V/t=1$.
\begin{figure}[ht]
\centerline{\includegraphics[width=8.5cm,height=5.0cm]{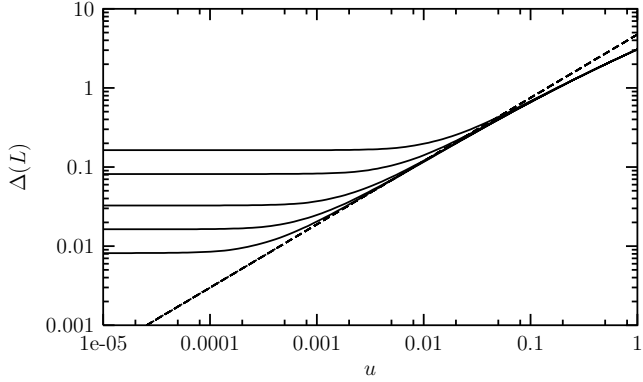}}
\caption{\label{udelta1} Energy gap $\Delta(L)$ (in units of $t$) 
as function of the dimerization parameter $u$
for $L=100, 200, 500, 1000, 2000$ (from top to bottom). The dashed line is the analytic
result of Eq.\ (\ref{eq:delta}) valid for $L \rightarrow \infty$ and $\Delta \ll \Delta_0$.}
\end{figure}

We now turn to the calculation of the Drude weight $D$ within the SCHA formalism.
The Drude weight or charge stiffness is the central quantity to characterize
charge transport at zero frequency.
At $T = 0$ the real part of the electrical conductivity at
frequency $\omega$ is of the form $\sigma(\omega) = 2\pi D \delta(\omega)
+ \sigma_{\rm reg}(\omega)$ with $\sigma_{\rm  reg}(\omega)\rightarrow 0$ for
$\omega\rightarrow 0$ in a system without impurities.
Therefore $D = 0$ characterizes an
insulator while $D > 0$ describes an (ideal) conductor.
The simplest way to calculate $D$ is due to
Kohn \cite{Kohn64} who has shown that the Drude weight can be
expressed as
\begin{equation}
\label{eq:Kohn}
D = \frac{L}{2} \left.\frac{{\rm d}^2 E}{{\rm d}\varphi^2}\right|_{\varphi=0}
\end{equation}
where $E(\varphi)$ is the ground state energy of
a ring of circumference $L$ which is threaded by the flux $\varphi$.
Alternatively, the parameter $\varphi$ can also be associated with a change of
boundary conditions, i.e. $\varphi = 0$ corresponds to periodic and
$\varphi = \pi$ to antiperiodic boundary conditions.
In the fermionic model (\ref{eq:dim-term})
the hopping parameter $t$ picks up a phase factor $\exp(\pm i\varphi)$
when a flux $\varphi$ is applied. In the bosonized form (\ref{eq:totalhamilt})
the only modification is that the current operator $J$ has to be replaced by
\cite{Loss92}
$J + \frac{\varphi}{\pi}$.
Correspondingly, we modify the $B$-dependent part of the
the trial Hamiltonian and write
\begin{equation}
\label{eq:HtrialB,phi}
 H_{\rm tr}^B(\varphi)=-iB(F^+_L F_R - F^+_R F_L) +
   \frac{\pi v g}{2L} \left({J}+\frac{\varphi}{\pi}\right)^2
\end{equation}
Applying the same procedure as in the case $\varphi = 0$
yields a variational estimate $\tilde E$ for the ground state energy
which now depends on the flux $\varphi$, i.e.\ $\tilde E =
\tilde E(\Delta,B,\varphi)$ where $\Delta$ and $B$, obtained
from the gap equations (\ref{eq:varpar}), are also
functions of $\varphi$.
On first sight it might seem impossible to calculate the Drude weight using
Eq.\ (\ref{eq:Kohn}) since there is no analytical solution of the
gap equations (\ref{eq:varpar}) even for $\varphi = 0$. However, a closer
look reveals a great deal of simplification.
From the second derivative

\begin{eqnarray}
\nonumber
\frac{{\rm d}^2 \tilde E}{{\rm d}\varphi^2}  & = &
\frac{\partial^2 \tilde E}{\partial \Delta^2}
\left(\frac{\partial \Delta}{\partial \varphi}\right)^2
+ \frac{\partial^2 \tilde E}{\partial B^2}
\left(\frac{\partial B}{\partial \varphi}\right)^2
+ \frac{\partial \tilde E}{\partial \Delta}
       \frac{\partial^2 \Delta}{\partial \varphi^2}\\
& &
+ \frac{\partial \tilde E}{\partial B}
\frac{\partial^2 B}{\partial \varphi^2}
+ \frac{\partial^2 \tilde E}{\partial \varphi^2}
+ \makebox{mixed terms}
\end{eqnarray}
one retains only
\begin{equation}
\left.\frac{{\rm d}^2 \tilde E}{{\rm d}\varphi^2}\right|_{\varphi=0} =
\left.\frac{\partial^2 \tilde E(\Delta,B,\varphi)}{\partial \varphi^2}\right|_{\varphi=0}
\end{equation}
since {\it i)} $\frac{\partial \tilde E}{\partial \Delta} =
\frac{\partial \tilde E}{\partial B} = 0$ due to the minimum
condition of the energy and {\it ii)}
$\frac{\partial \Delta}{\partial \varphi} = \frac{\partial B}{\partial \varphi} = 0$
at $\varphi=0$ due to symmetry
($\Delta$ and $B$ are even functions of $\varphi)$.
Since the trial Hamiltonian $H_{\rm tr} = H_{\rm tr}^\Delta + H_{\rm tr}^B(\varphi)$
consists of two commuting parts we may write
$\tilde E(\Delta,B,\varphi) = \tilde E(\Delta) + E_0(B,\varphi)$
where $\tilde E(\Delta)$ depends only on $\Delta$
and $E_0(B,\varphi)$ is the ground state energy of $H_{\rm tr}^B(\varphi)$
i.e.\ we obtain the simple result
\begin{equation}
D = \frac{L}{2}
\left.\frac{\partial^2 E_0(B,\varphi)}{\partial\varphi^2}\right|_{\varphi=0}
\end{equation}
where $B$ is given by Eq. (\ref{eq:varpar}).
To proceed we again represent $H_{\rm tr}^B(\varphi)$
in terms of the Mathieu equation (\ref{eq:HB}) where now
the boundary conditions are $\Psi(p_J+\pi)=e^{i\varphi}\Psi(p_J)$.
It is now straightforward to calculate the Drude weight
for a finite system of size $L$ in the gapped phase $g < 2$.
In the ``finite size gap'' region ($L \Delta \ll v$) we obtain
\begin{equation}
\label{Drude1}
D = D_0 (1 - \frac{q^2}{2} + \ldots)
\end{equation}
where $ q = 2LB / \pi v g \propto u L^{2-g}$ and $D_0 = vg / 2\pi$ is
the Drude weight of an unperturbed Luttinger liquid.
In the opposite limit ($L \Delta \gg v$) which corresponds to
a Mathieu equation with a large cosine potential the variation of the
ground state energy with change of boundary conditions is exponentially small
(as expected from the WKB approximation). A more careful treatment \cite{Abramowitz} yields
\begin{equation}
\label{Drude2}
 D \simeq D_0 \frac{4 }{\pi} \left(\frac{L \Delta}{vg}\right)^{3/2}
    {\rm e}^{-{2 L \Delta}/{\pi vg}}
\end{equation}
where $\Delta$ and $u$ are related via Eq.\ (\ref{eq:delta}).
In the special case of free fermions i.e.\ for $g=1$ and $v=2t$
we may compare the results obtained within the
SCHA with the exact Drude weight (see Appendix)
\begin{equation}
\label{Drudeex}
 D_{\rm ex}\simeq D_0 \left(\frac{\pi L\Delta_{\rm ex}}{v}\right)^{1/2}
   {\rm e}^{-{L \Delta_{\rm ex}}/{2 v}}
.\end{equation}
Although the expected exponential behavior is recovered in the SCHA result,
the different numerical factor in the exponent shows the limitations of the method.

\section{Summary}

We have constructed an extension of the SCHA in order to account for the existence
of Klein factors in bosonized Hamiltonians with non-linear perturbations.
As an application we have investigated a model of spinless fermions with modulated
hopping. For the infinite system, both the value of the Luttinger parameter $g_c = 2$ where
the transition from a gapless to a gapped phase takes places
and the exponent $1/(2-g)$ that characterizes the opening of the gap for $u \rightarrow 0$ are correctly
obtained within the SCHA. However, the bending of the phase boundary for finite values of
$u$ is not reproduced. When considering a finite system
Klein factors cannot be ignored {\it a priori}.
Within our approach it turns out,
that the crossover region from a finite size gap to a true dimerization gap coincides
with the crossover to the region where the Klein factors become relevant.

The Drude weight reflects the sensitivity of the system
with respect to a change of boundary conditions and is related to the properties of
the current operator $J$ in the bosonized version of the Hamiltonian.
In a finite system with an energy gap $\Delta$ the Drude weight is expected to be nonzero
but exponentially small, $D \sim \exp(- {\rm const} \cdot L \Delta)$.
Our extended version of the SCHA allows to calculate the Drude weight in the insulating phase
and we confirm the exponential behavior.

Using the same concepts the method can also be applied
to study finite size effects in more complex models with nontrivial phase diagrams.

\section*{Acknowledgements}
We thank K.\ Sch\"onhammer and C.\ Schuster for helpful discussions.
We acknowledge financial support from the Deutsche Forschungsgemeinschaft (SPP 1073).

\begin{appendix}
\section{}

For $V = 0$ the ground state energy of Hamiltonian (1) for a system
of $N = 4M + 2$ lattice sites is given by
\begin{equation}
\label{aenergy}
E_0(\varphi) = - 2t\sum_{n=-M}^M \sqrt{\cos^2 k_n + u^2 \sin^2 k_n}
\end{equation}
where $k_n = (2 \pi n + \varphi)/L$ and the lattice constant is set to one. We introduce $z_n = {\rm e}^{ik_n}$
and the function $f(z) = 1/(z^L - {\rm e}^{i\varphi})$
which has single poles at $z_n$ with residua ${\rm Res} f(z)|_{z=z_n}  = z_n {\rm e}^{-i\varphi} /L$.
Expressing Eq. (\ref{aenergy}) in terms of $z_n$ we may replace the sum by
a contour integral and obtain
\begin{equation}
E_0(\varphi) = - \frac{tL {\rm e}^{i\varphi}\sqrt{1 - u^2}}{2} \oint_{\cal C} \frac{{\rm d}z}{2\pi i} \;
\frac{\sqrt{z^2 + z^{-2} + 2\gamma}}{z(z^L - {\rm e}^{-i\varphi})}
\end{equation}
where $\gamma = (1 + u^2)/(1 - u^2)$ and ${\cal C}$ is a contour that encloses
the singularities of $f(z)$. We chose ${\cal C}$ to be composed of two circles
with radii slightly larger or smaller than one, respectively. Substituting
$z \rightarrow 1/z$ the integral along the inner circle can be mapped
on the integral along the outer circle.
The square root in the numerator has
branch cuts along the imaginary axis in the interval
$[-y_1,y_1]$ and for $|y| > y_0$ where $\pm i y_0$ and $\pm i y_1$ are
the zeroes of the function under the square root.
Now we deform the integration contour along the branch cut from $iy_0$ to $i \infty$
and obtain for the energy difference $\Delta E = E_0(\pi) - E_0(0)$
\begin{equation}
  \Delta E  = \frac{4tL\sqrt{1 - u^2}}{\pi}
  \int_{y_0}^\infty \; \frac{{\rm d}y}{y}
\frac{\sqrt{y^2 + y^{-2} - 2\gamma}}{y^{L} - y^{-L}}
\end{equation}
Substituting $y = {\rm e}^x$ yields
\begin{equation}
\Delta E = \frac{4tL\sqrt{1 - u^2}}{\pi}  \int_{x_0}^\infty {\rm{d}}x \
\frac{\sqrt{2\cosh (2x) - 2\gamma}}{{\rm e}^{Lx} -{\rm e}^{-Lx}}
\end{equation}
with $x_0 = {\rm Arcosh} \gamma = \ln((1+u)/(1-u))$. For $Lx_0 \gg 1$ the integral is rapidly
cut off by the exponential and we may expand the square root around $x = x_0$.
The Drude weight is then
\begin{equation}
D = \frac{L \Delta E}{4} \simeq t \sqrt{\frac{2uL}{\pi}}
 {\rm exp}\left({-\frac{L}{2} \ln\frac{1+u}{1-u}} \right)
\end{equation}
Expanding the logarithm for $u \ll 1$ and inserting $\Delta_{\rm ex} = 4t u$
yields Eq.\ (\ref{Drudeex}).
\end{appendix}

\end{document}